\documentclass[10pt,conference]{IEEEtran}
\usepackage{cite}
\usepackage{amsmath,amssymb,amsfonts}
\usepackage{algorithm2e}
\usepackage{graphicx}
\usepackage{textcomp}
\usepackage{comment}
\usepackage{balance}

\def\BibTeX{{\rm B\kern-.05em{\sc i\kern-.025em b}\kern-.08em
    T\kern-.1667em\lower.7ex\hbox{E}\kern-.125emX}}

\usepackage{packages}

\begin{document}

\title{How Do Developers Maintain and Evolve Their Agents' Instructions? An Empirical Study}

\author{
\IEEEauthorblockN{
Gianmario Voria\IEEEauthorrefmark{1},
Alfonso Cannavale\IEEEauthorrefmark{1},
Andrea De Lucia\IEEEauthorrefmark{1},
Yutaro Kashiwa\IEEEauthorrefmark{2},
Gemma Catolino\IEEEauthorrefmark{1},
Fabio Palomba\IEEEauthorrefmark{1}
}
\IEEEauthorblockA{\IEEEauthorrefmark{1}University of Salerno, Italy}
\IEEEauthorblockA{\IEEEauthorrefmark{2}Nara Institute of Science and Technology, Japan}
}

\maketitle

\begin{abstract}
%\textit{Context.} Autonomous coding agents are increasingly used to support software development, shifting parts of the engineering process to AI assistance. While this automation brings clear benefits, it introduces challenges in terms of governance, traceability, and control over agent behavior. Agent Context Files (ACFs) have recently emerged as a practical mechanism to guide agents through structured instructions. However, little is known about how these artifacts are maintained in practice and how their evolution relates to code development.
%\textit{Objective.} This paper proposes to systematically investigate the evolution of ACFs and their role in agent-driven development. Specifically, we will (1) classify ACF changes through a taxonomy grounded in software maintenance theory, (2) analyze how different types of changes are associated with code quality outcomes, and (3) examine their temporal patterns across the development lifecycle.
%\textit{Method.} We will conduct a large-scale mining study combining repositories with ACFs and agent-generated commits. We will mine ACF evolution at the commit level, classify changes using a qualitative approach, and analyze their association with structural and process-level code quality metrics. Statistical analyses and hypothesis testing will be used to evaluate differences across maintenance categories. The results are expected to inform both researchers and practitioners on how to design, maintain, and use ACFs more effectively to govern autonomous coding agents.
\textit{Context.} Autonomous coding agents are increasingly used in software development, shifting parts of the engineering process to AI assistance. While this automation brings clear benefits, it introduces challenges in governance, traceability, and control over agent behavior. Agent Context Files (ACFs) have emerged as a practical mechanism to guide agents through structured instructions, yet little is known about how these artifacts are maintained and how their evolution relates to code development.
\textit{Objective.} This paper plans to investigate the evolution of ACFs and their role in agent-driven development. Specifically, we (1) classify ACF changes through a taxonomy grounded in software maintenance theory, (2) analyze how different types of changes are associated with code quality outcomes, and (3) examine their temporal patterns across the development lifecycle.
\textit{Method.} We conduct a large-scale mining study combining repositories with ACFs and agent-generated commits. We reconstruct ACF evolution at the commit level, classify changes using a qualitative approach, and analyze their association with code quality metrics. Statistical analyses and hypotheses are used to evaluate differences across maintenance categories, to inform future design of ACFs for governing autonomous coding agents.
\end{abstract}

\begin{IEEEkeywords}
Coding Agents; Software Maintenance and Evolution; Empirical Software Engineering; Agent README.
\end{IEEEkeywords}

\section{Introduction}
Autonomous coding agents are reshaping software engineering (SE) processes \cite{hassan2025agentic}. While practitioners traditionally managed the entire lifecycle, recent AI advances have shifted part of this responsibility to autonomous agents capable of executing complex tasks with minimal human intervention \cite{li2025aiteammates}.
%Autonomous coding agents are reshaping the way software engineering (SE) processes are conceived and executed \cite{hassan2025agentic}. Traditionally, SE practitioners have been responsible for the design, development, evaluation, and maintenance of software systems. However, recent advances in AI have progressively shifted part of this responsibility toward autonomous agents, which are now capable---when appropriately instructed---of performing complex tasks with limited human intervention \cite{li2025aiteammates}.

While automation has long been a central objective of SE research, historically focused on supporting developers in specific tasks, the current paradigm represents a substantial shift. Rather than assisting isolated activities, autonomous coding agents can orchestrate entire segments of the development lifecycle, introducing a new level of autonomy in software production \cite{hou2024large}. This transition brings clear benefits in terms of efficiency and scalability, but also introduces new challenges related to control, transparency, and governance \cite{roychoudhury2025agentic}.

As development and evaluation activities become increasingly automated, greater attention must be devoted to upstream and downstream phases, particularly design and maintenance \cite{Gao2024The}. For instance, in requirements engineering, developers must now formulate specifications that are not only correct but also sufficiently precise and structured to guide autonomous agents, often in the absence of subsequent human interpretation or validation \cite{Lu2025Requirements}. Similarly, in maintenance, the coexistence of human- and agent-generated code introduces new complexities, including difficulties in tracing authorship and intent, understanding design decisions, and managing novel artifacts such as prompts and agent configurations \cite{Borg2025Echoes}.

% Among these emerging artifacts, a recent and increasingly adopted practice in agent-enabled software development is the use of
One increasingly adopted artifact in this space is the \textit{Agent Context File (ACF)} \cite{chatlatanagulchai2025use}. These artifacts can be seen as structured specifications that define how agents should behave within a project, including guidelines, constraints, and interaction rules. As such, ACFs act as an interface between human intent and agent execution, and have been proposed as a mechanism to reintroduce a degree of human governance over autonomous development processes \cite{chatlatanagulchai2025agent}.

While ACFs hold promise as governance artifacts, our understanding of how they are used and evolve in practice remains limited. \revised{In particular, while prior work has examined the effect of prompts and contextual information on agent behavior \cite{mohsenimofidi2025context, le2025impacts, 10992485}, considerably less attention has been devoted to understanding how these persistent governance artifacts are maintained over time and how their evolution relates to the evolution of the codebase.}
% Studying the evolution of ACFs is important for multiple reasons. First, it provides insights into the challenges practitioners face when instructing and controlling autonomous agents. For example, frequent or substantial changes to ACFs may indicate misalignment between intended and actual agent behavior, highlighting limitations in how instructions are specified or interpreted. Second, analyzing how ACF changes relate to code evolution can reveal patterns of effective and ineffective practices, offering empirical guidance for designing and maintaining these artifacts. Finally, understanding when and how often ACFs are updated can shed light on the dynamics of how developers maintain and adapt governance mechanisms throughout the development lifecycle.

Studying the evolution of ACFs is important for several reasons. First, it provides insights into the challenges of instructing and controlling autonomous agents, as frequent or substantial changes may indicate misalignment between intended and actual behavior. Second, relating ACF changes to code evolution can reveal effective and ineffective practices, offering guidance for their design and maintenance. Finally, understanding when and how often ACFs are updated sheds light on how developers adapt governance mechanisms throughout the development lifecycle.

%\textit{Accordingly, the objective of this work is to investigate how ACFs are maintained and evolve in software repositories and to understand the role of such evolution in shaping subsequent development activities.} Rather than solely describing how ACFs change over time, we adopt a theoretically grounded perspective based on classical software maintenance theory \cite{swanson1976dimensions, mockus2000identifying}. Specifically, we investigate whether changes to ACFs conform to established maintenance categories (i.e., corrective, adaptive, and perfective) and use this taxonomy as an analytical lens to study how different types of ACF modifications relate to code quality outcomes and exhibit distinct temporal patterns throughout the development lifecycle.
\vspace{0.2cm}
\steResearchQuestionBox{\faLightbulbO \hspace{0.05cm} \textbf{Main Hypothesis.}
We hypothesize that Agent Context Files act as evolving governance artifacts in agent-enabled software development, and that their modifications are not arbitrary but systematically reflect developers' attempts to control and adapt agent behavior. In particular, we expect that ACF changes conform to classical software maintenance categories, and that different types of changes are associated with distinct code quality outcomes and exhibit different temporal and frequency patterns throughout the development lifecycle.}
\vspace{0.2cm}

To this end, we plan a large-scale mining study of repositories that adopt ACFs. We will extract commit-level information on ACF modifications and code changes, reconstruct their evolution over time, and \revised{derive an ACF-specific taxonomy of change types and interpret the resulting categories through the lens of classical software maintenance theory.} We will then analyze these changes along three dimensions: (i) their distribution across categories, to assess the applicability of classical maintenance theory; (ii) \revised{their association with code quality outcomes through both structural metrics (complexity, coupling, size) and process-level indicators of corrective activity}; and (iii) their temporal and frequency patterns within the development lifecycle. %Through this statistically grounded analysis, we aim to provide a comprehensive understanding of how ACFs are maintained and how they influence agent-driven development.

Beyond characterizing ACF evolution, our study is designed to yield actionable \textbf{implications} for both researchers and practitioners. \revised{For researchers, our taxonomy will provide a theoretically grounded framework for future studies on governance artifacts in agentic settings, while our findings on code quality outcomes will contribute empirical evidence on how the maintenance and evolution of persistent governance artifacts relate to agent-generated development outcomes. }\revised{For practitioners, the findings can support the development of evidence-based governance practices for autonomous coding agents by identifying which types of ACF modifications are most common, when they are typically introduced, and how they relate to subsequent development outcomes.}%For practitioners, the temporal and quality-related patterns we will uncover can directly inform the design of evolution guidelines for ACFs: concrete insights on when to update governance artifacts, what types of changes are most beneficial, and how to avoid governance drift throughout the development lifecycle.

\section{Background and Related Work}

% ============================================================
\smallskip
\textbf{Autonomous Coding Agents in Software Engineering.}
% ============================================================
The role of AI in software engineering has shifted from single-turn code completion toward fully autonomous agents that decompose high-level goals into sub-tasks, execute multi-step plans, and interact with tools such as test runners, debuggers, and version control systems with minimal human intervention~\cite{wang2025agents, hassan2025agentic}. Wang et al.~\cite{wang2025agents} characterize these systems through a perception--memory--action framework that highlights a form of explicit procedural memory encoded in the agent's configuration, which governs how the agent perceives its environment and selects actions. This shift has enabled agents to take on increasingly complex SE activities, including automated program repair~\cite{bouzenia2025repairagent}, autonomous testing~\cite{feldt2023towards}, and agentic refactoring~\cite{horikawa2025agentic}.
Greater autonomy, however, introduces governance challenges: empirical work has documented integration defects and security risks in agentic settings~\cite{shao2024llms}, and studies consistently identify the absence of authoritative project-specific knowledge as a primary failure mode~\cite{wang2025agents}.
% Greater autonomy, however, comes with new governance challenges. Empirical work has documented integration defects and security risks that arise specifically in agentic settings~\cite{shao2024llms}, and studies consistently note that effective deployment depends not only on model capability but on how well agents are configured to understand project-specific conventions, constraints, and architectural decisions~\cite{wang2025agents, hassan2025agentic}. The absence of such grounding is identified as a primary failure mode, leading to outputs that may be locally correct but misaligned with team practices~\cite{wang2025agents}.

% ============================================================
\smallskip
\textbf{Software Maintenance Theory.}
% ============================================================
We ground our ACF evolution analysis in software maintenance theory (SWEBOK v4.0~\cite{washizaki2024guide}), which distinguishes between corrections (corrective, preventive, emergency) and enhancements (adaptive, perfective, additive).
%which distinguishes two macro-categories. \textit{Corrections} aim to preserve operability and include \textit{corrective} (reactive fixes after failures), \textit{preventive} (proactive fixes of latent faults), and \textit{emergency} maintenance (urgent, temporary interventions). \textit{Enhancements} aim to evolve the system and include \textit{adaptive} (responding to environmental changes), \textit{perfective} (improving quality attributes), and \textit{additive} maintenance (introducing new functionality).  \textit{Adaptive} maintenance, which responds to environmental changes, can fall under either macro-category depending on whether the adaptation is necessary to preserve correct functioning (correction) or represents an improvement (enhancement).
The operationalization of such classifications on version control data has been empirically demonstrated by Trautsch et al.~\cite{trautsch2023really}, providing methodological grounding for our approach.

\smallskip
\textbf{Empirical Studies on Agent Context Files.}
To address the governance gap described above, practitioners have adopted a class of structured Markdown files that encode persistent, project-level instructions for autonomous agents.
These artifacts---named \texttt{CLAUDE.md}, \texttt{AGENTS.md}, or \texttt{copilot-instructions.md} depending on the tool---act as explicit procedural memory that developers write to shape agent behavior within a project, a mechanism that prior work has shown to affect generated code quality~\cite{shrivastava2023repository, le2025impacts}.

Chatlatanagulchai et al.~\cite{chatlatanagulchai2025use, chatlatanagulchai2025agent} analyzed over 2,300 ACFs from 1,925 repositories, finding that these files are actively maintained through small incremental edits focused on functional concerns, with non-functional requirements rarely specified. Mohsenimofidi et al.~\cite{mohsenimofidi2025context} report no established structural standard across the 466 projects they examined. These studies confirm the practical relevance of ACFs but treat their evolution only marginally, without characterizing what changes are made, how they affect the codebase, or when they occur.

\smallskip
\textbf{Code Quality of AI-Generated Code.}
A growing body of empirical work has begun to examine the quality of code produced by LLMs and autonomous agents. Jamil et al.~\cite{jamil2025can} compared GPT-3.5-Turbo and GPT-4 outputs against human-written solutions, finding that LLMs, when guided by advanced prompts, can outperform humans on several metrics.

Studies using SonarQube to assess LLM-generated code have consistently found that functional correctness does not predict overall code quality~\cite{sabra2025assessing}: even code that passes all unit tests still carries latent bugs, security vulnerabilities, and code smells. In agentic settings specifically, Horikawa et al.~\cite{horikawa2025agentic} found that agent-generated refactoring yields small but significant improvements in structural code metrics. %Chatlatanagulchai et al.~\cite{chatlatanagulchai2025agent} complement this by observing that developers rarely include non-functional guidelines such as security and performance in their ACFs. 
% Whether systematic changes to ACF instructions are associated with measurable differences in downstream code quality has not been studied.

\smallskip
\textbf{Commit-Level Code Quality Measurement.}
% A complementary body of work has examined how code quality metrics evolve at the commit level. Trautsch et al.~\cite{trautsch2023really} analyzed changes in static metrics across 54 open-source Java projects, classifying commits as perfective or corrective based on Swanson’s taxonomy~\cite{swanson1976dimensions}. They found that perfective commits consistently reduce cyclomatic complexity, coupling, and code size, whereas corrective commits tend to increase these metrics. We adopt their delta-based approach—computing metric differences between a commit and its parent—and apply it to agent-generated commits following ACF changes, using their findings to inform our hypotheses for RQ2.
Trautsch et al.~\cite{trautsch2023really} analyzed static metric changes across 54 open-source Java projects, classifying commits as perfective or corrective per Swanson's taxonomy~\cite{swanson1976dimensions}. They found that perfective commits consistently reduce cyclomatic complexity, coupling, and code size, whereas corrective commits tend to increase them. We adopt their delta-based approach---computing metric differences between a commit and its parent---and apply it to agent-generated commits following ACF changes, using their findings to inform our hypotheses for RQ2.

% At the process level, Amit and Feitelson~\cite{amit2021corrective} introduced the Corrective Commit Probability (CCP), defined as the proportion of corrective commits within a development period. Validated on 7,557 GitHub projects using a linguistic model on commit messages (93\% accuracy), CCP has been shown to correlate with larger files, higher coupling, and lower productivity, making it a reliable, language-agnostic proxy for code quality. We adapt CCP to the window level, computing it between consecutive ACF changes.

At the process level, Amit and Feitelson~\cite{amit2021corrective} proposed the Corrective Commit Probability (CCP), defined as the share of corrective commits within a development period. Validated on 7,557 GitHub projects with 93\% accuracy, CCP correlates with larger files, higher coupling, and lower productivity, providing a reliable and language-agnostic proxy for code quality. We adapt CCP to the window level by computing it between consecutive ACF changes.

\section{Empirical Study Design}\label{sec:design}

\revised{Our goal is to study how ACFs are maintained and evolve by deriving a taxonomy of their changes and analyzing their association with code quality and temporal patterns.}

\subsection{Research Questions}
%\topar{\yutaro{Can we structure RQs more systematically? For example, when, what, who, why, etc.? With the current RQs, reviewers might wonder why these research questions were made. }}
%We will structure our empirical study around three research questions (\textbf{RQ\textsubscript{s}}).

While ACFs are increasingly adopted to guide autonomous coding agents, their evolution in practice remains poorly understood. Prior work has largely treated them as static specifications~\cite{chatlatanagulchai2025use, chatlatanagulchai2025agent}, overlooking how developers iteratively refine these artifacts over time. % Simply observing that ACFs change, however, is insufficient to derive actionable insights, motivating the need for theoretically grounded analyses.
In our \textbf{first RQ}, we will address this gap by grounding the analysis in software maintenance theory \cite{washizaki2024guide}, treating ACF modifications as instances of established maintenance categories. \revised{Specifically, we will derive an ACF-specific taxonomy of change categories through qualitative analysis and subsequently examine how the resulting categories relate to established software maintenance concepts.}

\steResearchQuestionBox{\textbf{RQ\textsubscript{1}}. \emph{To what extent can changes to Agent Context Files be characterized according to classical software maintenance categories, and how are they distributed?}}

While characterizing ACF changes is a necessary first step, it is not sufficient to understand their role in the development process. As artifacts designed to guide autonomous coding agents, ACFs may shape how code is generated and evolves. Consequently, their modifications may alter agent behavior and be associated with measurable differences in the resulting code.
Different types of ACF changes may reflect distinct intents and correspond to different outcomes in the codebase. Understanding these relationships is essential to assess the practical impact of ACFs, particularly in terms of code quality.

To this end, in our \textbf{second RQ}, we will investigate whether different types of ACF modifications are associated with different code quality outcomes in subsequent code evolution, providing insights to inform their design and maintenance.

%While characterizing ACF changes is a necessary first step, it is not sufficient to understand their role in the development process. \textit{ACFs are designed to guide the behavior of autonomous coding agents and, as such, may shape how code is generated and evolves over time.} Consequently, modifications to these artifacts may alter the instructions provided to the agent, potentially leading to measurable differences in the resulting code.

%In this context, different types of ACF changes may reflect distinct intents and may be associated with different outcomes in the codebase. Understanding whether and how these categories of change relate to subsequent development activities is essential for assessing the practical impact of ACFs, particularly in terms of code quality. 

%To this end, in our \textbf{second RQ}, we investigate whether different types of ACF modifications exhibit differences in their association with code quality outcomes in subsequent code evolution. Such insights can inform both researchers and practitioners on how to design and maintain ACFs more effectively.

\steResearchQuestionBox{\textbf{RQ\textsubscript{2}}. \emph{Do different types of Agent Context File changes exhibit differences in their association with code quality outcomes in subsequent code evolution?}}

While identifying the types of ACF changes and their association with code quality outcomes provides insight into what these changes are and what they do, it does not fully capture how they unfold throughout the development process. In particular, the timing and frequency of ACF modifications may reflect different maintenance intents and strategies adopted by developers when interacting with autonomous agents.

From a software maintenance perspective, different categories of changes are expected to exhibit distinct temporal dynamics. For example, corrective changes may occur reactively in response to failures or undesired agent behavior, whereas perfective changes may be introduced proactively to refine or improve instructions \cite{washizaki2024guide}. Similarly, update frequency may indicate how actively developers adjust governance mechanisms over time.
Understanding whether different types of ACF changes exhibit distinct temporal and frequency patterns is therefore essential to characterize how these artifacts are maintained, which will be the aim of our \textbf{third RQ}. %These insights can inform guidelines on when and how to update ACFs effectively.

%While identifying the types of ACF changes and their association with code quality outcomes provides insight into what these changes are and what they do, it does not fully capture how they unfold throughout the development process. In particular, the timing and frequency of ACF modifications may reflect different maintenance intents and strategies adopted by developers when interacting with autonomous agents.

%From a software maintenance perspective, different categories of changes are expected to exhibit distinct temporal dynamics. For instance, corrective changes may occur reactively in response to failures or undesired agent behavior, whereas perfective changes may be introduced more proactively to refine or improve existing instructions \cite{swanson1976dimensions}. Similarly, the frequency of updates may indicate how actively developers adjust governance mechanisms over time.

%Understanding whether different types of ACF changes exhibit distinct temporal and frequency patterns is therefore essential to characterize how these artifacts are maintained in practice, which is the aim of our \textbf{third RQ}. Such insights can provide a deeper understanding of the lifecycle of ACFs and inform guidelines on when and how to update them effectively.

\steResearchQuestionBox{\textbf{RQ\textsubscript{3}}. \emph{Do different types of Agent Context File changes exhibit distinct temporal and frequency patterns throughout the development lifecycle?}}

%\tofigure{\gianmario{method figure summarizing the mining process and the RQs answer}}

\subsection{Experimental Setup}
\label{sec:setup}
%\topar{\gianmario{\color{red}{@Alfonso, I sketched up this section in simple terms haha. If you want we can discuss about the RWs and metrics to measure, otherwise feel free to decide (given the difference in timezone it could be difficult) and then when we meet all together we talk about it!}}}
\smallskip
%\textbf{Datasets.} \gianmario{Here we should describe the two datasets used for mining. The first one is AIDev, which contains PRs related to agent generated code. From this, we have to explicitly say that we trust their way to classify agent generated code, which we use too. Secondo one is the dataset of repos with ACFs, should explain its content and again state that we reuse their classification methods where needed.}
\textbf{Datasets.} 
We will rely on two complementary datasets to study the relationship between ACFs and software evolution.

First, we will use the \emph{AIDev} dataset \cite{li2025aiteammates}, which provides a curated collection of repositories (\textit{116,211}) and pull requests (\textit{932,791}) involving agent-generated code. Agent contributions are identified through a validated classification pipeline, which we will adopt to extract commit-level information (e.g., file changes, diffs, metadata) for analyzing development activity.

Second, we will use an existing dataset of repositories containing ACFs \cite{chatlatanagulchai2025use}, comprising \textit{2,303} context files across \textit{1,925} repositories, including their content and versioning history. ACFs are identified through established naming conventions (e.g., \texttt{AGENTS.md}, \texttt{CLAUDE.md}, \texttt{copilot-instructions.md}) and represent persistent configuration artifacts used to guide agent behavior. We will reuse the original classification procedures where applicable.

 \revised{The study feasibility is supported by the original ACF dataset \cite{chatlatanagulchai2025use}, containing over 10,000 ACF-modifying commits (for \textbf{RQ\textsubscript{1}}). For the remaining analyses, a preliminary pipeline extracted 10,763 commit snapshots with context files, 18,213 commits with file metadata, and 8,600 commits with intersecting information, confirming that sufficient data are available to conduct the study.}
 %The final sample will depend on the intersection of the two datasets and filtering criteria (e.g., availability of commit history, presence of ACFs changes, completeness of metadata).

\smallskip
\textbf{Variables and Measures.} 
%\gianmario{here we have to outline all the variables considered in our RQs (basically the metrics). How to call this part? what are these?}
%The study uses two units of analysis: ACF-modifying commits for RQ\textsubscript{1} and RQ\textsubscript{3}, and development windows between consecutive ACF-modifying commits for RQ\textsubscript{2}. In particular, we consider (i) commits that modify ACFs, representing units of ACF change, and (ii) subsequent commits in the same repository, representing development outcomes potentially associated with such changes. Importantly, \emph{we consider as related all agent-generated code commits occurring between an ACF-modifying commit and the immediately following ACF-modifying commit, regardless of category}. The category of the ACF change is then used as the independent variable to group and compare the associated code outcomes.
%\emph{we consider changes to be related to an ACF change from that same ACF modifying commit to the next ACF modifying commit that falls into a different category.}
The study will use two units of analysis: ACF-modifying commits (for RQ\textsubscript{1} and RQ\textsubscript{3}) and development windows between consecutive ACF-modifying commits (for RQ\textsubscript{2}). Each ACF-modifying commit defines a window that includes all subsequent agent-generated code commits up to the next ACF change. These commits will be treated as outcomes associated with the preceding ACF modification. The ACF change category, once classified, will be used as the independent variable for all the comparisons.

To address \textbf{RQ\textsubscript{1}}, we will characterize ACF modifications according to software maintenance categories (\emph{corrections} and \emph{enhancements})\cite{washizaki2024guide} and, more specifically, their subcategories (\textit{corrective}, \textit{preventive}, \textit{adaptive}, \textit{perfective}, and \textit{additive}). This variable captures the nature of the change applied to the ACF and is used to analyze the distribution of maintenance categories and assess whether ACF evolution conforms to established maintenance patterns.

To address \textbf{RQ\textsubscript{2}}, we will measure code quality outcomes along two complementary dimensions. The first captures \textit{structural code quality} through the delta of static metrics between each agent-generated commit and its parent, specifically cyclomatic complexity, lines of code, and coupling~\cite{trautsch2023really, mccabe1976complexity}. \revised{We draw on Trautsch et al.~\cite{trautsch2023really} as a methodological precedent for measuring software quality evolution through changes in quality metrics. Unlike their study, which investigates the direct impact of maintenance activities on code quality, our analysis examines the association between ACF modifications and the quality of subsequent agent-generated code.} Structural metrics are computed at the commit level and then aggregated at the window level. The second captures \textit{process-level quality} through the Corrective Commit Probability (CCP)~\cite{amit2021corrective}, computed over the window of agent-generated commits between two consecutive ACF-modifying commits. Corrective commits are identified via a linguistic model applied to commit messages, making this measure fully language-agnostic and applicable across the heterogeneous repositories in our dataset.

\begin{comment}
To address \textbf{RQ\textsubscript{2}}, we measure code quality outcomes using a set of established, language-agnostic metrics commonly adopted in mining software repositories studies~\cite{bavota2015experimental}. Specifically, we consider:

\begin{itemize}[leftmargin=*]
    \item \textit{Code churn metrics}, including lines of code added and deleted, and the number of modified files, as proxies for change impact and development effort \cite{farago2015cumulative, purushothaman2005toward};
    \item \textit{Complexity metrics}, such as cyclomatic complexity, as indicators of code maintainability and understandability \cite{haraldsson2025aspects, curtis2006measuring};
    \item \textit{Structural metrics}, including the number of functions or classes affected by a change, to capture the structural scope of modifications \cite{Kondo2019The, Kouroshfar2013Studying}.
\end{itemize}

These measures are computed at the commit level and capture properties of the code produced or modified following ACF changes. In this analysis, the type of ACF change is used to group observations and assess whether different categories are associated with different code quality outcomes. We intentionally adopt language-agnostic measures to ensure generalizability across heterogeneous repositories.
\end{comment}

To address \textbf{RQ\textsubscript{3}}, we will define variables capturing the timing and frequency of ACF changes throughout the development lifecycle. Specifically, we will consider \textit{relative timing}, computed as the normalized position of a commit within the repository history, and \textit{change frequency}, measured as the number of ACF modifications over a given number of commits or time window. These measures will allow us to analyze whether different types of ACF changes exhibit distinct temporal and frequency patterns.

To account for potential \textbf{confounding effects}, we will consider additional variables at the commit, file, and repository level, including commit or ACF size, repository size, and activity level. \revised{Programming language will be recorded as a potential confounding variable and, where sample sizes permit, stratified analyses will be performed on the most represented languages.} These variables will be included to isolate the relationship between ACF changes and the observed outcomes.

%\gianmario{for RQ0 is quite easy, maybe here we could just discuss the maintenance evolution theory that we will use as base for the taxonomy building. The actual metric is distribution of each type of change, so its just quick explanation. Since we will use this distribution in our hypotheses (maybe X type of changes more prevalent than Y) should write something for the statistical part?}
%\gianmario{for RQ1 is quite hard. Here we have to explain the kind of quality metrics we want to use. I was thinking that we may want to select (1) code (cyclomatic?) complexity, (2) LOC, (3) lack of cohesion (lcom)??. But in geneeral we have to find a related work who does so and try to comply with their quality metrics. Another point is: do we want to consider code smells? If so, we would have to reduce the analysis to python/java but that may basically reduce the scope too much (we may have too few files to evaluate in commits for only 2 langs).}
%\gianmario{for RQ3 we have to select the moment of the repo lifetime that we want to consider (we should select based on some RW), and i dont know if we need metric for frequency? maybe is not a variable}

\subsection{Data Collection}
%\topar{\gianmario{very low level from the analyses I have already done, should be reviewed!}}
To investigate ACF evolution in actual code repositories, we will collect and combine commits from two publicly available datasets as described in Section~\ref{sec:setup}.
%: (i) the AIDev dataset of agent-authored commits~\cite{li2025aiteammates} and (ii) a curated dataset of repositories containing agent context files~\cite{chatlatanagulchai2025agent}.

%The first step of the data collection process will involve a \textit{mining} phase. Particularly, to investigate ACF evolution in actual code repositories, we will collect and combine commits from two publicly available datasets as described in Section \ref{sec:setup}: (i) the AIDev dataset of agent-authored commits \cite{li2025aiteammates} and (ii) a curated dataset of repositories containing agent context files \cite{chatlatanagulchai2025agent}. 

\smallskip
\textbf{ACFs Mining.} We use the ACF dataset to identify repositories and commits that modify primary context files \cite{chatlatanagulchai2025agent}. \revised{In this study, the term ACF refers exclusively to the primary repository-level context files identified in the original dataset (e.g., \texttt{CLAUDE.md}, \texttt{AGENTS.md}, and \texttt{copilot-instructions.md}). Auxiliary artifacts such as skills, imported context fragments, or task-specific instruction files are outside the scope of the analysis.} We will consider all repositories in this dataset and treat each commit modifying a context file as a unit of change.
We extract repository-level metadata and static representations of the context files, including identifiers (owner, name, URL), temporal metadata (creation date, first adoption), structural properties (e.g., length and complexity), and full textual content.
To support longitudinal analysis, we will reconstruct ACF evolution by tracking commits that modify context files. For each, we will extract both the current and parent versions of the ACF, capturing modifications as before/after states. We also record commit-level traces, including timestamps and change metrics (e.g., lines and structural changes), enabling fine-grained analysis of ACF evolution.
%\textbf{ACFs Mining.} We will start from an existing dataset of agent context files, collected by mining GitHub repositories for files following the official naming conventions of widely used autonomous coding agents \cite{chatlatanagulchai2025agent}. Specifically, we will consider all the repositories made available in this dataset. ACF modifications will be treated at the commit level, with each commit that affects a context file representing a unit of change for subsequent analysis. 

%We will extract repository-level metadata and static representations of the context files, including repository identifiers (owner, name, URL), temporal metadata (creation date, first context file adoption), structural properties (e.g., length and complexity), and full textual content. 

%To support longitudinal analysis, we will reconstruct the evolution of ACFs by identifying commits that modify context files through file-level change tracking. For each commit affecting an ACF, we will extract both the version of the context file at the commit and its version in the immediate parent revision. This will allow us to explicitly capture ACF modifications as before/after states, enabling fine-grained analysis of how these artifacts evolve over time. Additionally, we will record commit-level traces of such modifications, including timestamps and change metrics (e.g., lines added/removed and structural changes).

\smallskip
\textbf{Agentic Repositories Mining.} 
We will integrate the ACF dataset with the AIDev dataset \cite{li2025aiteammates}, which provides a large-scale collection of agent-authored pull requests and commits. Since ACFs are available only for Claude Code, OpenAI Codex, and GitHub Copilot, we will restrict the analysis to these agents and retain only commits from repositories where a context file is present. Integration will be done at the repository level using canonical identifiers (owner/repository). %Although this limits the analysis to projects that both adopt ACFs and contain detectable agent-generated contributions, it ensures ecological validity by focusing on settings where explicit governance mechanisms are in place.

From the filtered AIDev data, we will construct a commit-level dataset capturing structural and semantic properties of agent-generated changes. For each commit, we consider metadata (identifier, message, author, timestamp), size metrics (additions, deletions, total changes), file-level information (filenames and programming languages), and task-related annotations where available. Temporal information (e.g., timestamps and ordering) will be used to derive timing and frequency patterns of ACF updates throughout the development lifecycle.

Additionally, we will implement the refined classification procedure proposed by Li et al. \cite{li2025aiteammates} to identify commits generated by specific agents. This procedure combines multiple signals, including author identifiers, commit message patterns, and naming conventions. While this approach provides a best-effort approximation, we acknowledge that agent attribution may not always be perfectly accurate.

\smallskip
\textbf{Commit Reconstruction.}
For each commit in the filtered dataset, we reconstruct a detailed snapshot of the development state. Specifically, we extract the full commit metadata and patch representation, enabling line-level analysis of changes. In addition, we  retrieve all modified files in both the commit version and the immediate parent revision.

% For commits that modify the agent context file, we will explicitly extract both the version of the ACF at the commit and its version in the immediate parent revision. This will allow us to represent ACF modifications as before/after states, enabling fine-grained analysis of the changes applied to these artifacts. Additionally, 
For every commit, regardless of whether the ACF is modified, we will retrieve the version of the agent context file present in the repository at that specific point in time by querying the repository tree at the corresponding revision. This ensures that, for each commit, we can reconstruct the exact instructions available to the agent when producing changes.

To support the measurement of structural code quality for RQ\textsubscript{2}, the extracted diff and pre- and post-change file versions will be used to compute the delta in static metrics between each agent-generated commit and its parent. These per-commit deltas will then be aggregated over development windows defined by consecutive ACF-modifying commits.

\smallskip
\textbf{Data Quality and Filtering.}
We will include all commits from the filtered repositories that are associated with the selected agents. Repositories that are empty, inaccessible, or lack sufficient commit history will be excluded. A repository will be considered as adopting ACFs if at least one valid ACF is identified in its history and thus has been included in the original dataset \cite{chatlatanagulchai2025agent}. For each commit, the presence of an ACF will be determined at the specific revision under analysis. Commits occurring before the introduction of an ACF in the repository will be retained but excluded from the main analyses, as no active governance artifact was in place at the time of their production. To ensure data quality, we will remove duplicated or malformed entries and verify the consistency of repository identifiers across datasets. \textbf{All data and scripts will be made publicly available.}

\begin{comment}
\smallskip
\textbf{Resulting Mined Data.}
The final dataset will consist of a collection of commits enriched with:
\begin{itemize}
    \item repository-level context (including the presence and content of agent context files),
    \item commit-level metadata and structural properties,
    \item explicit representations of ACF modifications (captured as before/after versions of the context files),
    \item and fine-grained code artifacts capturing both the pre-change and post-change state of the modified files.
\end{itemize}
\end{comment}

\subsection{Data Analysis}
\textbf{To answer RQ\textsubscript{1}}, we will analyze all commits that modify an ACF to derive a taxonomy of ACF changes. \revised{Following Ralph's recommendations for taxonomy development in qualitative software engineering research \cite{ralph}, we will employ an inductive thematic analysis approach, as our goal is to identify and characterize recurring patterns of ACF evolution rather than develop a novel explanatory theory.}

We will first draw a statistically representative random sample of ACF-changing commits from the mined dataset. For each sampled commit, annotators will inspect: (i) the commit message, (ii) the ACF diff, (iii) the previous and current version of the ACF, and, when needed, (iv) the surrounding code changes in the same commit. The goal of this inspection is to identify the intent of the ACF modification.

\revised{The analysis will proceed iteratively. Annotators will assign short descriptive codes to each ACF change (e.g., ``clarify testing command'' or  ``add coding convention''), reusing existing codes when appropriate and creating new ones when necessary. Through constant comparison and discussion, the codebook will be progressively refined and consolidated. Semantically related codes will then be grouped into higher-level themes representing recurring categories of ACF evolution. This process will continue until the taxonomy stabilizes and no substantial new categories emerge from the analyzed data.} \revised{Because a single ACF modification may contain multiple logically distinct changes (e.g., removing obsolete instructions while simultaneously introducing new ones), annotators will assign multiple codes to the same modification if necessary.}

Each sampled change will be labeled independently by at least two annotators. \revised{Inter-rater agreement will be measured using Cohen's $\kappa$. We will consider $\kappa \geq 0.70$ as the minimum acceptable agreement level before proceeding with large-scale classification. If agreement falls below this threshold, the coders will refine the codebook and repeat the calibration process. To enhance rigor and trustworthiness, we will maintain an audit trail documenting coding decisions and taxonomy refinements throughout the analysis process.} This process follows taxonomy-construction guidelines in software engineering \cite{Usman2017Taxonomies}, which require specifying the subject, descriptive basis, procedure, and validation strategy; here, these correspond to ACF changes, commit/ACF diffs, qualitative coding, and agreement-based validation.

\revised{Once the initial taxonomy has been established, we will perform a deductive interpretation step to relate the identified categories to established software maintenance and evolution concepts \cite{washizaki2024guide}. Specifically, we will analyze how the emergent ACF change categories align with maintenance activities such as corrective, preventive, adaptive, perfective, and additive maintenance. The purpose of this step is not to guide the coding process, but rather to position the resulting taxonomy within the broader software engineering literature and enable comparison with prior studies on software evolution.}

After defining the entire taxonomy on the sampled data, we will classify the remaining ACF-changing commits. Depending on the dataset size, this step can be performed either manually or through an assisted procedure. For large datasets, we will adopt an LLM-as-judge approach, prompting the model with taxonomy definitions and labeled examples, and validating the resulting labels on an additional randomly sampled subset, following approaches used in prior studies on ACF content~\cite{chatlatanagulchai2025agent}.
\revised{The emergent ACF-specific categories constitute the primary outcome of the qualitative analysis. The maintenance-theory labels derived through the deductive interpretation step will subsequently be used as the analytical categories in RQ\textsubscript{2} and RQ\textsubscript{3}, enabling hypothesis testing, comparison across repositories, and alignment with established software evolution theory.}

To analyze the distribution of ACF changes across categories, we will follow statistical practices from empirical software engineering. As such data are often non-normal and heterogeneous~\cite{Kitchenham2017Robust}, and RQ\textsubscript{1} involves categorical counts, we will use a chi-square goodness-of-fit test to assess whether changes are uniformly distributed across categories \cite{Bentler1980Significance}. We will complement significance testing with effect size reporting to quantify the magnitude of observed differences~\cite{Kitchenham2017Robust}.

To analyze the distribution of ACF changes, we will test two complementary null hypotheses addressing the distribution of changes across (a) macro-categories and (b) subcategories. 

\faLightbulbO \ \textbf{\underline{H1\textsubscript{a}}} \textit{There is no significant difference in the distribution of ACF changes classified as corrections versus enhancements.}

\faLightbulbO \ \textbf{\underline{H1\textsubscript{b}}} 
\textit{There is no significant difference in the distribution of ACF changes across maintenance subcategories (corrective, preventive, adaptive-correction, adaptive-enhancement, perfective, and additive).}

If H1\textsubscript{b} is rejected, we will conduct post-hoc pairwise comparisons using proportion tests with Holm-Bonferroni correction. Rejecting any of these hypotheses would indicate that ACF evolution is not uniformly distributed and that certain maintenance types are more prevalent than others.

%\faLightbulbO \ \textbf{\underline{H1\textsubscript{b}}} \textit{There is no significant difference in the distribution of ACF changes between corrective and preventive maintenance.}

%\faLightbulbO \ \textbf{\underline{H1\textsubscript{c}}} \textit{There is no significant difference in the distribution of ACF changes across adaptive, perfective, and additive maintenance.}

%If H1\textsubscript{b} or H1\textsubscript{c} are rejected, we will conduct post-hoc pairwise comparisons using proportion tests with Holm-Bonferroni correction within the respective macro-category. Rejecting any of these hypotheses would indicate that ACF evolution is not uniformly distributed and that certain maintenance types are more prevalent than others.

%If H1\textsubscript{a} is rejected, we will conduct post-hoc pairwise comparisons using proportion tests with Holm-Bonferroni correction to control the family-wise error rate, testing the following pairwise null hypotheses:

% \faLightbulbO \ \textbf{\underline{H1\textsubscript{b-c-d}}} \textit{There is no significant difference in the proportion of changes between any pair of maintenance categories.}

% Rejecting these hypotheses would indicate that ACF evolution is not uniformly distributed and that certain types of maintenance categories are more prevalent than others.

\smallskip
\textbf{In RQ\textsubscript{2}}, we plan to analyze whether different ACF changes are associated with differences in agent-generated code quality.

For each commit $N$ that modifies an ACF, we will consider the interval of commits between $N$ and the next ACF-modifying commit $M$ in the same repository as the ACF window. \revised{The window is therefore defined by ACF snapshots rather than temporal proximity. Autonomous coding agents interact with the current version of the ACF available in the repository regardless of when that version was last modified. Consequently, all agent-generated commits occurring between two consecutive ACF modifications are considered to operate under the same governance configuration and are associated with the same ACF snapshot. }For the last ACF-modifying commit in a repository, the window extends to the last commit in the mined history.  Each window will be associated with the category of the ACF change introduced in commit $N$, classified according to the taxonomy defined in RQ\textsubscript{1}. Within each window, we will extract all agent-generated commits and exclude non-code artifacts. 
For each commit within the window, we will compute the delta of structural code quality metrics between the modified files and their previous versions. These metrics will then be aggregated at the window level (mean and median) to obtain a single representation of structural quality change for each ACF modification. In addition, we will compute the Corrective Commit Probability (CCP) over the same window. This captures process-level quality in terms of rework and defect-related activity.

%We will compare quality measures across maintenance categories using non-parametric tests, as quality metrics are not expected to follow normal distribution~\cite{Kitchenham2017Robust}. Specifically, we will apply the Kruskal–Wallis test to assess overall differences. When significant, we will perform post-hoc pairwise comparisons using the Wilcoxon rank-sum test with Holm-Bonferroni correction. We will report statistical significance and effect sizes, using Cliff’s Delta for pairwise comparisons. We define the following null hypotheses:
To evaluate code quality across maintenance categories, we will employ the non-parametric Kruskal-Wallis test, followed by post-hoc pairwise Wilcoxon rank-sum tests with Holm-Bonferroni correction \cite{Kitchenham2017Robust}. Effect sizes will be quantified via Cliff’s Delta. We test the following null hypotheses:

\faLightbulbO \ \textbf{\underline{H2\textsubscript{a}}} \textit{There is no significant difference in code quality outcomes between ACF maintenance subcategories.}
% \faLightbulbO \ \textbf{\underline{H2\textsubscript{a}}} \textit{There is no significant difference in code quality outcomes between ACF changes classified as corrections and those classified as enhancements.}

If H2\textsubscript{a} is rejected, we will test pairwise hypotheses:

\faLightbulbO \ \textbf{\underline{H2\textsubscript{b-c-d}}} \textit{There is no significant difference in quality outcomes between any pair of maintenance subcategories.}

\smallskip
\textbf{In RQ\textsubscript{3}}, we will analyze when ACF changes occur during the development lifecycle and how frequently different types of changes are applied. We will define the lifecycle as the interval from the later of (i) the first detected ACF introduction and (ii) the first agent-generated commit, to the last mined commit. Within this interval,  we will identify all ACF-modifying commits and compute their \revised{\emph{relative timing}, operationalized through two complementary measures: (i) commit-relative timing, computed as the normalized position of the commit within the repository history, and (ii) time-relative timing, computed as the normalized position of the commit within the agent-enabled lifecycle in terms of elapsed time.}
%as the normalized position in the repository history (commit index over total commits), enabling comparison across repositories of different sizes and durations. 
We will measure the frequency of ACF changes for each maintenance category along two dimensions: the number of modifications per fixed number of commits (e.g., per 100 commits) and per unit of time (e.g., per day), capturing activity-relative and temporal update intensity. These measures will be computed per maintenance category to compare their temporal and frequency patterns.

%We will compare these measures across categories using non-parametric tests~\cite{Kitchenham2017Robust}. Specifically, we will apply the Kruskal–Wallis test for overall differences and, when significant, perform post-hoc pairwise comparisons using the Wilcoxon rank-sum test with Holm-Bonferroni correction. We will report statistical significance and effect sizes, using Cliff’s Delta for pairwise comparisons. 
For statistical analysis, we will use the same procedure as in \textbf{RQ\textsubscript{2}}. We define the following null hypotheses:

\faLightbulbO \ \textbf{\underline{H3\textsubscript{a}}} \textit{There is no significant difference in the temporal distributions of ACF changes across maintenance subcategories.}

\faLightbulbO \ \textbf{\underline{H4\textsubscript{a}}} \textit{There is no significant difference in the frequency distributions of ACF changes across maintenance subcategories.}

If H3\textsubscript{a} or H4\textsubscript{a} is rejected, we will test the following pairwise hypotheses:

\faLightbulbO \ \textbf{\underline{H3\textsubscript{b}}} \textit{There is no significant difference in timing between any pair of maintenance subcategories.}

\faLightbulbO \ \textbf{\underline{H4\textsubscript{b}}} \textit{There is no significant difference in frequency between any pair of maintenance subcategories.}

\section{Threats to Validity}
This section outlines potential threats to the validity of our study and the mitigation strategies that will be applied.

\textbf{Internal validity} concerns whether observed relationships can be attributed to the studied factors. A key threat is linking code quality outcomes to ACF changes, as ACFs contain heterogeneous instructions and modifications may affect only parts of the file, limiting causal interpretation. To mitigate this, we adopt a window-based design and control for confounders (commit size, ACF size, repository size, activity level), framing results as associations rather than causation. Another threat concerns identifying agent-generated commits, which relies on heuristics; we mitigate this using validated procedures \cite{li2025aiteammates}. \revised{Lastly, because ACFs govern heterogeneous aspects of development, some categories of modifications may have only indirect or limited influence on the selected quality measures, potentially reducing the strength of observable associations.}

\textbf{Construct validity} concerns how well measures capture the intended concepts. Code quality is multifaceted, and our use of structural and process-level metrics captures only part of it. We mitigate this by using complementary, literature-grounded metrics and prioritizing language-agnostic measures. \revised{Another threat concerns the operationalization of temporal evolution. Relative commit position may not perfectly reflect lifecycle stage, so we employ two complementary timing measures: a commit-based relative position and a time-based relative position computed over the agent-enabled lifecycle. Differences in agent performance across programming languages may influence the observed outcomes. We mitigate this by explicitly recording language information and performing language-stratified analyses where feasible. Finally, classifying ACF changes may introduce subjectivity, as the taxonomy is derived through inductive qualitative analysis. Following established recommendations for rigorous qualitative software engineering research \cite{ralph}, we mitigate this threat through independent coding by multiple researchers, iterative codebook refinement, inter-rater agreement assessment, disagreement resolution procedures, and the maintenance of an audit trail documenting coding decisions.}

\textbf{External validity} concerns generalizability. Our study relies on publicly available repositories that adopt ACFs and autonomous coding agents \cite{chatlatanagulchai2025use}, which may not fully represent industrial or proprietary settings. Moreover, the analysis is limited to agents for which both ACFs and agent-generated commits are available (Codex, Copilot, Claude).

\textbf{Conclusion validity} concerns the reliability of statistical inferences. To mitigate statistical threats, we adopt established practices in empirical software engineering \cite{Kitchenham2017Robust}, including non-parametric tests, effect size reporting, and corrections for multiple comparisons, and we predefine hypotheses and analyses to ensure reproducibility.
\section{Conclusion}
%This report proposes an empirical study of how Agent Context Files (ACFs), which act as governance mechanisms for autonomous coding agents, are maintained and evolve.

%Our contribution is threefold. First, we assess whether ACF changes conform to classical software maintenance categories and derive a taxonomy grounded in established theory. Second, we analyze whether different types of ACF modifications are associated with variations in the quality of agent-generated code. Third, we examine the timing of these changes throughout the development lifecycle. Our findings are expected to inform both research and practice. For researchers, the taxonomy and empirical evidence provide a foundation for future work on governance in agentic settings. For practitioners, the results offer actionable guidance on when and how to evolve ACFs to effectively guide autonomous coding agents.
%Our findings are expected to inform both research and practice by offering empirical insights into how governance artifacts can be effectively designed and maintained to guide autonomous coding agents.

This report proposes a mining study on the maintenance and evolution of ACFs as governance artifacts for autonomous coding agents. By deriving an ACF-specific taxonomy and analyzing its relationship with code quality and temporal patterns, our findings are expected to inform both research and practice. For researchers, the taxonomy and empirical evidence provide a foundation for future work on governance in agentic settings. For practitioners, the results offer actionable guidance on when and how to evolve ACFs to guide autonomous coding agents effectively.

\begin{comment}
\section*{Data Availability}
The data that support the findings of this study are openly available in our online appendix \cite{appendix}.
\end{comment}

\section*{Acknowledgment}
We gratefully acknowledge the financial support of JSPS KAKENHI (JP24K02921, JP25K03100), JST ASPIRE (JPMJAP2415), and JST CREST (JPMJCR23M1, JPMJCR26X7). We also acknowledge the use of GPT to ensure linguistic accuracy and readability.
% Note: Yutaro could shorten the funding list

\balance
\bibliographystyle{IEEEtran}
\bibliography{references}

\end{document}